\documentclass[twocolumn,showpacs,amsmath,nofootinbib,pra,aps,amssymb,longbibliography]{revtex4-1}

\usepackage[T1]{fontenc}
\usepackage[utf8]{inputenc}
\usepackage{times}
\usepackage{color} 
\usepackage{array}
\usepackage{amssymb,amsmath}
\usepackage{amsbsy}
\usepackage[pdftex]{graphicx} 
\usepackage{bm}
\usepackage{float}
\usepackage{dcolumn}
 
\usepackage[unicode,breaklinks]{hyperref}
\hypersetup{
    unicode=true,
    plainpages=false, 
    colorlinks=true,
    linkcolor=blue,
    citecolor=blue,
    filecolor=black,
    urlcolor=blue
}
\usepackage{url}
\usepackage{verbatim}

\newcommand{\gdd}{g_\mathrm{dd}}
\newcommand{\add}{a_\mathrm{dd}}
\newcommand{\edd}{\epsilon_\mathrm{dd}}
\newcommand{\br}{\mathbf{r}}
\newcommand{\bx}{\mathbf{x}}

\newcommand{\Phidd}{\Phi_\mathrm{dd}}
\newcommand{\UD}{U_\mathrm{dd}}
\newcommand\gammaQF{\gamma_\mathrm{QF}}

\begin{document}

\title{Compressibility and speeds of sound across the superfluid to supersolid phase transition of an elongated dipolar gas} 
	\author{P.~B.~Blakie,$^{1,2}$  L.~Chomaz,$^{3}$ D.~Baillie,$^{1,2}$ and F.~Ferlaino$^{4,5}$  }
	
	\affiliation{%
	$^1$Dodd-Walls Centre for Photonic and Quantum Technologies, New Zealand\\
	$^2$Department of Physics, University of Otago, Dunedin 9016, New Zealand\\
	$^{3}$Physikalisches Institut der Universit\"at Heidelberg, Im Neuenheimer Feld 226, 69120 Heidelberg, Germany\\
	$^{4}$Institut f\"ur Experimentalphysik, Universit\"at Innsbruck, Technikerstra{\ss}e 25, 6020 Innsbruck, Austria\\
	$^{5}$Institut f\"ur Quantenoptik und Quanteninformation, \"Osterreichische Akademie der Wissenschaften, Technikerstra{\ss}e 21a, 6020 Innsbruck, Austria}	 
\date{\today} 
\begin{abstract}    
We investigate the excitation spectrum and compressibility of a dipolar Bose-Einstein condensate in an infinite tube potential in the parameter regime where the transition between superfluid and supersolid phases occurs. 
Our study focuses on the density range in which crystalline order develops continuously across the transition. 
Above the transition the superfluid shows a single gapless excitation band, phononic at small momenta and with a roton at a finite momentum. Below the transition, two gapless excitations branches (three at the transition point) emerge in the supersolid. 
We examine the two gapless excitation bands and their associated speeds of sound in the supersolid phase. Our results show that the speeds of sound and the compressibility are discontinuous at the transition, indicating a second-order phase transition. These results provide valuable insights into the identification of supersolid phenomena in dipolar quantum gases and the relationship to supersolidity in spin-orbit coupled gases.
\end{abstract} 

\maketitle

\section{Introduction}
Experiments with dipolar Bose-Einstein condensates (BECs) have observed the transition to a supersolid ground state  \cite{Tanzi2019a,Bottcher2019a,Chomaz2019a}, and have studied its elementary excitations \cite{Tanzi2019b,Guo2019a,Natale2019a}. The supersolid state breaks translational invariance by developing a spatially modulated (crystalline) structure, while still exhibiting superfluidity. For a  $D$-dimensional crystal the supersolid state will exhibit $D+1$ gapless excitation branches. These reflect the number of Nambu-Goldstone modes associated with the spontaneously broken symmetries of the supersolid state \cite{Watanabe2012a}. The excitations can be classified by the character of fluctuations they cause  \cite{Macri2013a,Ancilotto2013a,Natale2019a}.
 Although there is hybridization of the properties of the branches, $D$ of the gapless branches are generally associated with density fluctuations and are termed density or phonon branches. The remaining gapless branch of excitations is associated with phase fluctuations and is referred to as a phase or Bogoliubov branch, related to superfluid aspects of the system (i.e.~tunneling of atoms between unit cells).

\begin{figure}[htb!]  
   \includegraphics[width=3.1in]{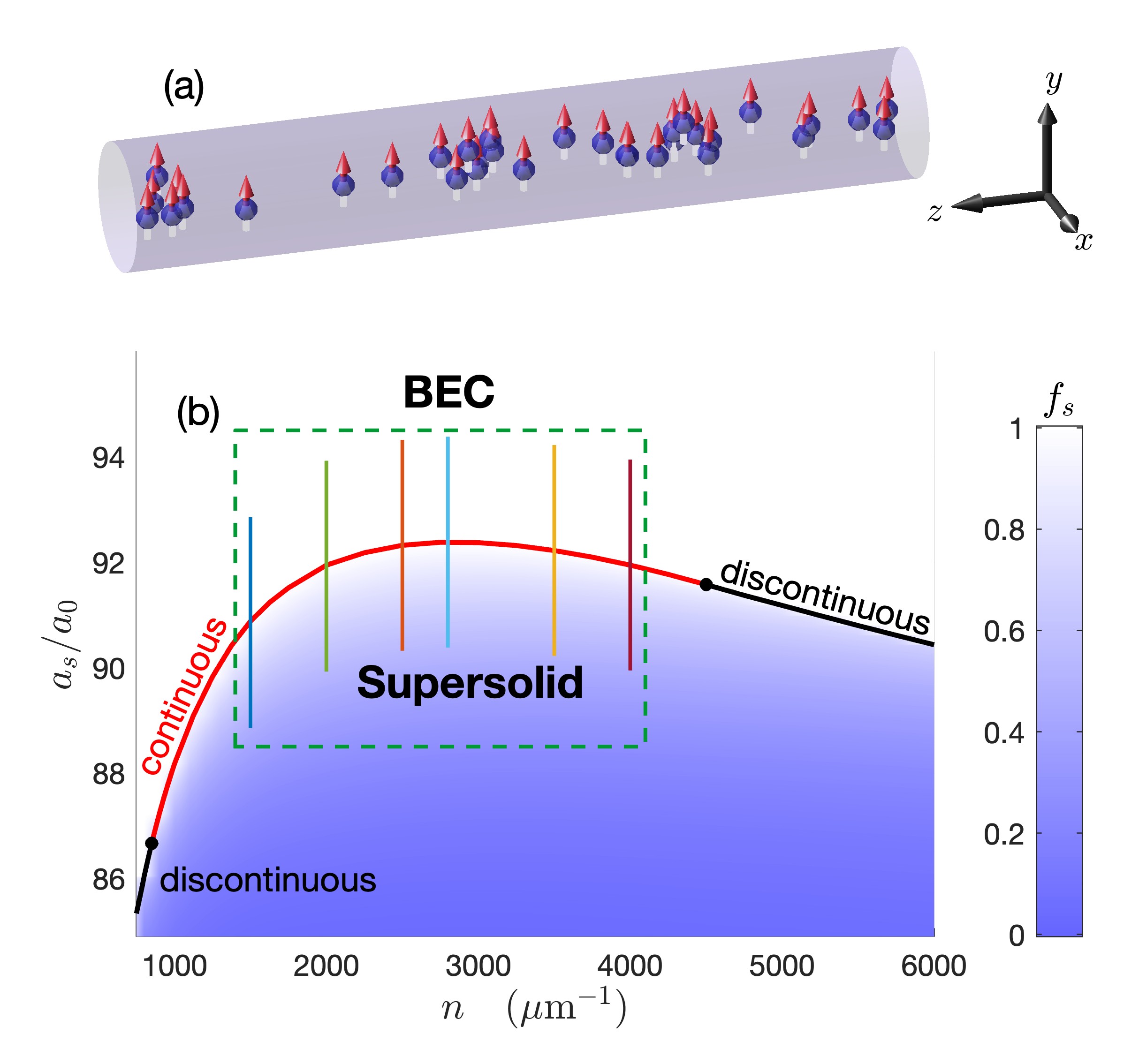} 
   \caption{(a) Schematic figure of thermodynamic system:  infinite tube-shaped potential confining dipoles polarized along $y$. (b) Phase diagram showing the superfluid fraction of the ground state as a function of density $n$ and the s-wave scattering length $a_s$. The uniform BEC (superfluid) and modulated supersolid states are shown separated by a continuous (red line) or discontinuous (black line) transition.   Dashed box indicates the parameter regime we consider in this work, focusing on the continuous transition. Vertical colored lines show the parameter regime for the transition data considered in Fig.~\ref{fig:compress}(a).
Phase diagram for  $^{164}$Dy using $a_{\mathrm{dd}}=130.8\,a_0$  and with $\omega_{\rho}=2\pi\times150\,$Hz (see \cite{Smith2023a}).
   \label{fig:schematic}}
\end{figure}

The majority of recent experimental work with dipolar BECs have used cigar shaped potentials in which a  $D=1$ supersolid transition can occur.  The relevant thermodynamic limit of this system is an infinitely long tube trap, i.e.~a system with transverse confinement only and a fixed linear density \cite{Roccuzzo2019a,Blakie2020a,Blakie2020b,Ilg2023a,Smith2023a}  [see Fig.~\ref{fig:schematic}(a)]. 
 It is found that, depending on the density, the supersolid transition in the thermodynamic system can be continuous or discontinuous \cite{Blakie2020b,Ilg2023a,Smith2023a}. The continuous transition emerges for a range of intermediate densities, and occurs when a roton excitation  gap of the uniform BEC state vanishes  \cite{Roccuzzo2019a,Ilg2023a,Blakie2020b} [see Fig.~\ref{fig:schematic}(b)]. Experimental and theoretical studies of the finite system also reveal similar behavior (e.g.~see \cite{Hertkorn2019a,Hertkorn2021a,Biagioni2022a}).
 It has also been shown that two gapless excitation branches  (a density and a phase  branch)  emerge in the transition to the 1D supersolid state \cite{Roccuzzo2019a,Tanzi2019b,Guo2019a,Natale2019a,Petter2021a,Ilg2023a}. We note that in the finite sized experimental systems the supersolid transition is revealed by a bifurcation in the compressional excitations of the gas \cite{Tanzi2019b,Natale2019a}, which can be interpreted as in-phase and out-of-phase combinations of the gapless excitations of the thermodynamic system.

We also note studies of supersolid properties in the thermodynamic regime for bosons with dipole-dipole interactions (DDIs)  with $D=2$ \cite{Lu2015a,Zhang2019a} or general soft-core interactions with $D=1$ \cite{Sepulveda2008a}, $D=2$ \cite{Pomeau1994a,Saccani2012a,Hsueh2012a,Kunimi2012a,Macri2013a}, or $D=3$ \cite{Pomeau1994a,Henkel2010a,Ancilotto2013a}. For the $D=1$ soft-core system the transition is always continuous\footnote{The discontinuous regime for the $D=1$ tube confined dipolar BEC emerges from the three-dimensional character of the system}. For the $D>1$ cases the transition to the supersolid state is generally first order\footnote{There is a critical point for the $D=2$ DDI case where the transition is continuous at particular critical density \cite{Zhang2019a}.}  and the system properties change abruptly at the transition.   
 
Spin-orbit coupled BECs also exhibit a phase transition to a supersolid-like stripe phase, which has been observed in experiments \cite{Li2017a}. For this realization the coupling to an optical field produces a $D=1$  (supersolid) stripe phase (irrespective of system dimension) and two gapless excitation branches  are predicted \cite{Li2013a}. 
The transition from the uniform planewave phase to the stripe phase is first-order \cite{Li2012a,Li2015a}, in contrast to the $D=1$ soft-core case and the dipolar case at intermediate densities [Fig.~\ref{fig:schematic}(b)]. Features of the spin-orbit system, such as the speeds of sound and compressibility in the thermodynamic regime, have been theoretically studied across the transition \cite{Li2013a,Li2015a,Martone2021a}. These features remain unexplored for the  $D=1$ dipolar supersolid.  We address this here by examining the behaviour of the excitations, speeds of sound, density response, and compressibility of this system. We focus on the parameter regime where the crystalline order develops continuously [see Fig.~\ref{fig:schematic}(b)] and find that the compressibility and the speeds of sound change discontinuously across this transition. Thus indicating that the transition is second order.

  \section{System and ground states}
\begin{figure*}[htbp!] 
   \centering
   \includegraphics[width=7in]{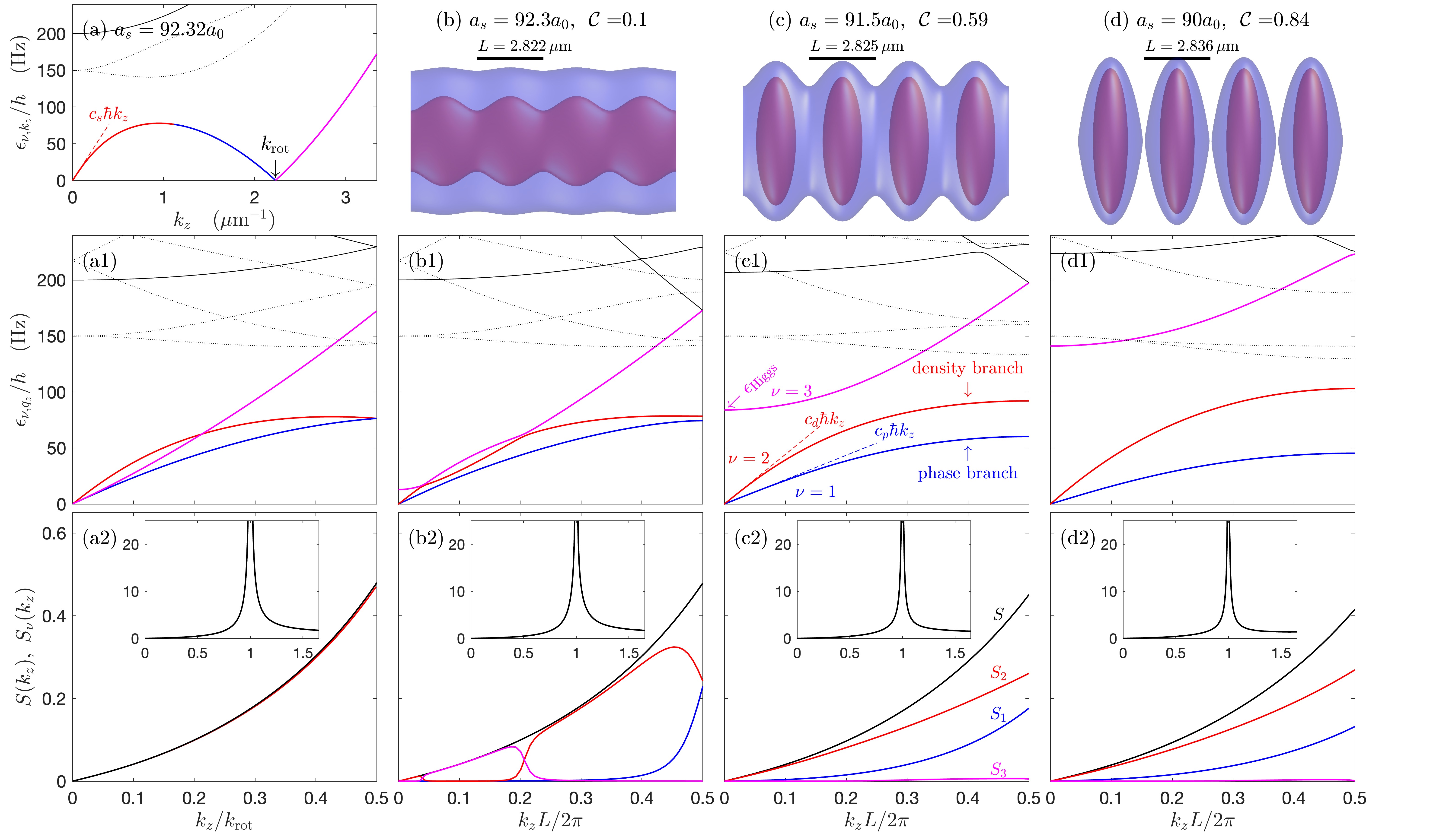}
   \caption{Density profiles, excitation spectra and structure factors of a dipolar Bose gas in an infinite tube. 
   (a) Excitation spectrum for a uniform system at $a_s=a_\mathrm{rot}$ where the roton softens. Excitations bands with even parity in $x$ and $y$ (solid lines), other bands (dotted lines). (a1) The results from (a) reduced to the first Brillouin zone with the colours of the lowest 3 even symmetry bands corresponding to the segments in (a). 
   (a2) The static structure factor $S(k_z)$ (black line) is dominantly contributed to by the lowest band  of (a) [i.e.~$S_1(k_z)$,   red line for the $k_z$ range considered]  (see Sec.~\ref{Sec:SFcompress}). Inset shows $S(k_z)$ over a wider momentum range. 
(b)-(d) Density isosurfaces of crystalline ground states for $a_s<a_{\mathrm{rot}}$. Red (blue) isosurface at $4\times10^{20}\,$m$^{-3}$ ($ 10^{20}\,$m$^{-3}$). Unit cell size $L$ and density contrast $\mathcal{C}$ are also indicated.
The excitation spectra (b1)-(d1)  and static structure factors  (b2)-(d2) corresponding to the  ground states in (b)-(d), using same line types as in (a1) and (a2), respectively.  The static structure factor is shown (black line) and individual contributions of the lowest three excitations bands (i.e.~$S_{\nu=1,2,3}$) [see labels in (c2)].
Results for  $^{164}$Dy using $a_{\mathrm{dd}}=130.8\,a_0$ for a linear density of $n=2500\,\mu$m$^{-1}$ and with $\omega_{\rho}=2\pi\times150$Hz.
   }
   \label{fig:dispersion}
\end{figure*}

Here we consider a gas of magnetic bosonic atoms in a radially symmetric tube potential $V=\frac{1}{2}m\omega_\rho(x^2+y^2)$, where $\omega_\rho$ is the angular trap frequency describing the transverse confinement. The theoretical description of this system is provided by an extended Gross-Pitaevskii equation (eGPE) which includes the leading order effects of quantum fluctuations  \cite{Ferrier-Barbut2016a,Chomaz2016a,Wachtler2016a,Bisset2016a}, and has been extensively used to model supersolid experiments with dipolar BECs (e.g.~see \cite{Tanzi2019a,Bottcher2019a,Chomaz2019a}).
The eGPE energy functional for this system is  
\begin{align}
E &= \int d\bx\, \psi^*\left[h_\mathrm{sp}+\tfrac12g_s|\psi|^2+\tfrac12\Phidd  +\tfrac25\gammaQF|\psi|^3\right]\psi,\label{Efunc}
\end{align}
where $h_\mathrm{sp}=-\frac{\hbar^2}{2m}\nabla^2+V$ is the single particle Hamiltonian. The short ranged interactions are governed by the coupling constant $g_s= 4\pi \hbar^2 a_s/m$ where $a_s$ is the $s$-wave scattering length.
The long-ranged DDIs are described by the potential
\begin{align}
\Phidd(\bx)=\int d\bx'\,\UD(\bx-\bx')|\psi(\bx')|^2,
\end{align}
where the atoms are polarized along $y$ with
$\UD(\br) = \frac{3\gdd}{4\pi r^3}\left(1-{3y^2}/{r^2}\right).$
Here $\gdd=4\pi\hbar^2\add/m$, with $\add = m\mu_0\mu_m^2/12\pi\hbar^2$ being the dipolar length, and $\mu_m$ the atomic magnetic moment. The effects of quantum fluctuations are described by the quintic nonlinearity with coefficient $\gammaQF = \frac{32}3 g_s\sqrt{a_s^3/\pi}\mathcal{Q}_5(\edd)$ where $\mathcal{Q}_5(x)=\Re\{\int_0^1 du[1+x(3u^2 - 1)]^{5/2}\}$ \cite{Lima2011a} and $\edd \equiv \add/a_{s}$. 

We constrain the stationary states of Eq.~(\ref{Efunc})  to have an average linear density of $n$, and ground states are found (following Ref.~\cite{Smith2023a}) by minimising the energy per particle.   For the case of modulated (crystalline) states, the system chooses a preferred unit cell size $L$, and the normalization constraint is $\int_{\mathrm{uc}}dz\int d\bm{\rho}\,|\psi|^2=nL$, where $\bm{\rho}=(x,y)$ represents the transverse coordinates, and  $\mathrm{uc}$ denotes the unit cell $z\in[-\tfrac12L,\tfrac12L]$.  These stationary states are  solutions of  the eGPE 
\begin{align}
\mu\psi=\left(h_\mathrm{sp}+ g_s|\psi|^2+\Phidd  + \gammaQF|\psi|^3 \right)\psi,
\end{align}
 where $\mu$ is the chemical potential.
 
 In Fig.~\ref{fig:dispersion} we present results illustrating the transition from a uniform to spatially modulated state as the $s$-wave scattering length is reduced. For the parameters considered in these results the ground state is uniform (superfluid state) for $a_s>a_{\mathrm{rot}}=92.32\,a_0$. Here $a_{\mathrm{rot}}$ is the value of the scattering length where a roton excitation goes soft as we will discuss in Sec.~\ref{Sec:excitations} (also see \cite{Roccuzzo2019a,Blakie2020a}). For  $a_s<a_\mathrm{rot}$ the ground state is modulated. We can characterise the strength of modulation using the density contrast 
 \begin{align}
 \mathcal{C}=\frac{n_{\max}-n_{\min}}{n_{\max}+n_{\min}},
 \end{align}
  where $n_{\max}$ and $n_{\min}$ are the maximum and minimum of the linear density $n(z)=\int d\bm{\rho}\,|\psi|^2$, respectively. Results for $\mathcal{C}$ show that  the density modulation develops continuously as $a_s$ decreases below $a_\mathrm{rot}$ [see Fig.~\ref{fig:dispersion}(b)-(d), Fig.~\ref{fig:sound}(a), and Refs.~\cite{Roccuzzo2019a,Blakie2020b,Ilg2023a,Smith2023a}].

\section{Excitations}\label{Sec:excitations}
The elementary quasi-particle excitations are described within the framework of Bogoliubov theory.  In this theory the excitations give the small deviations of the condensate field with respect to ground state as
\begin{align}
\Psi(\mathbf{x},t) =e^{-i\mu t/\hbar}&\left[\psi(\mathbf{x})  +\sum_{\nu,q_z}\left\{ c_{\nu,q_z}u_{\nu,q_z}(\mathbf{x})e^{-i\omega_{\nu,q_z}t} \right. \right. \nonumber \\
& \left. \left.  -c_{\nu,q_z}^*v^*_{\nu,q_z}(\mathbf{x})e^{i\omega_{\nu,q_z}^*t}\right\} \right].
\end{align}
Here  $\epsilon_{\nu,q_z}=\hbar\omega_{\nu,q_z}$ are the  quasi-particle energies, $\hbar q_z$ is a quasi-momentum in the first Brillouin zone, i.e.~$q_z\in[-\pi/L,\pi/L]$,  $\nu$ is the band index and  $c_{\nu,q_z}$ are the expansion amplitudes. The quasi-particle amplitudes take the Bloch form
\begin{align}
u_{\nu,q_z}(\mathbf{x})=\bar{u}_{\nu,q_z}(\mathbf{x})e^{iq_zz},\quad v_{\nu,q_z}(\mathbf{x})=\bar{v}_{\nu,q_z}(\mathbf{x})e^{iq_zz},
\end{align} 
where $\{\bar{u}_{\nu,q_z}(\mathbf{x}),\bar{v}_{\nu,q_z}(\mathbf{x})\}$ are periodic in the unit cell.
 More details of the Bogoliubov analysis of the eGPE can be found in Ref.~\cite{Baillie2017a}.

In Fig.~\ref{fig:dispersion} we show some examples of the excitation spectra of the system. The case in Fig.~\ref{fig:dispersion}(a) is for a uniform state at the transition point $a_s=a_\mathrm{rot}$. Here the $z$-momentum $\hbar k_z$ is a good quantum number for the excitations\footnote{In the uniform case we can  take $q_z=k_z$, due to translational invariance.}. A single Nambu-Goldstone branch exists in the uniform state, corresponding to the lowest energy excitation band that is gapless as $k_z\to0$. This reflects the broken gauge symmetry associated with superfluidity.
We observe a fully developed roton-like local minimum in the excitation spectrum that has softened to zero energy.  Here we define $a_{\mathrm{rot}}$ as the value of $a_s$ where the roton feature of the uniform state has  a minimum at zero-energy  (i.e.~a fully softened roton). We note that $a_\mathrm{rot}$ varies with  $n$ and confinement. The wavevector of the softened roton is denoted $k_\mathrm{rot}$ and the density modulation first develops at the transition point with a unit cell size set by the roton wavelength, i.e.~as $a_s\to a_{\mathrm{rot}}$ from below,  $L\to2\pi/k_{\mathrm{rot}}$.  To aid in our later comparison to the excitations of the modulated ground states, in Fig.~\ref{fig:dispersion}(a1) we map the uniform state excitation results of subplot (a) onto the positive part of first Brillouin zone ($q_z\in[-\frac{1}{2}k_{\mathrm{rot}},\frac{1}{2}k_{\mathrm{rot}}]$), taking $k_{\mathrm{rot}}$ as the reciprocal lattice vector. The color coding of segments of the lowest excitation band used in (a) is selected to help identify the features in the reduced zone scheme. In particular, we note that the soft roton feature manifests in the reduced zone representation as 2 additional gapless excitations bands (blue and magenta color).  

In Figs.~\ref{fig:dispersion}(b1)-(d1) we show the excitation spectra corresponding to the modulated ground states shown in (b)-(d), respectively.
The case in (b1) is close to the transition with a weak modulation. Here the spectrum is similar to the roton case [cf.~(a1)], but the ground state modulation causes some noticeable changes in the lowest three excitations bands:  a gap (for $q_z\to0$) develops in the $\nu=3$ (magenta color) band as we move away from the transition, while the other two bands remain gapless, and avoided crossings occur where the $\nu=2$ (red color) and $\nu=3$ (magenta color) bands approach each other.   For the more strongly modulated cases (c1) and (d1), the three lowest excitation bands separate, and can be unambiguously assigned. We denote the lowest gapless band (blue color) as the phase band, and the higher gapless band (red color) as the density band [see Fig.~\ref{fig:dispersion}(c1)].  The identification of these bands can be made by assessing the dominant effect of the excitations on the phase or density fluctuations of the system as has been done for dilute supersolid states with soft-core interactions and DDIs (e.g.~see \cite{Macri2013a,Ancilotto2013a,Natale2019a}). In the spin-orbit coupled stripe phase, similarly two gapless bands emerge, but these are identified as spin and density nature, due to their effect on density and spin fluctuations \cite{Li2013a}.

 \section{Superfluidity, characteristic excitations and speeds of sound} 

\begin{figure}[htb!]  
   \includegraphics[width=3.3in]{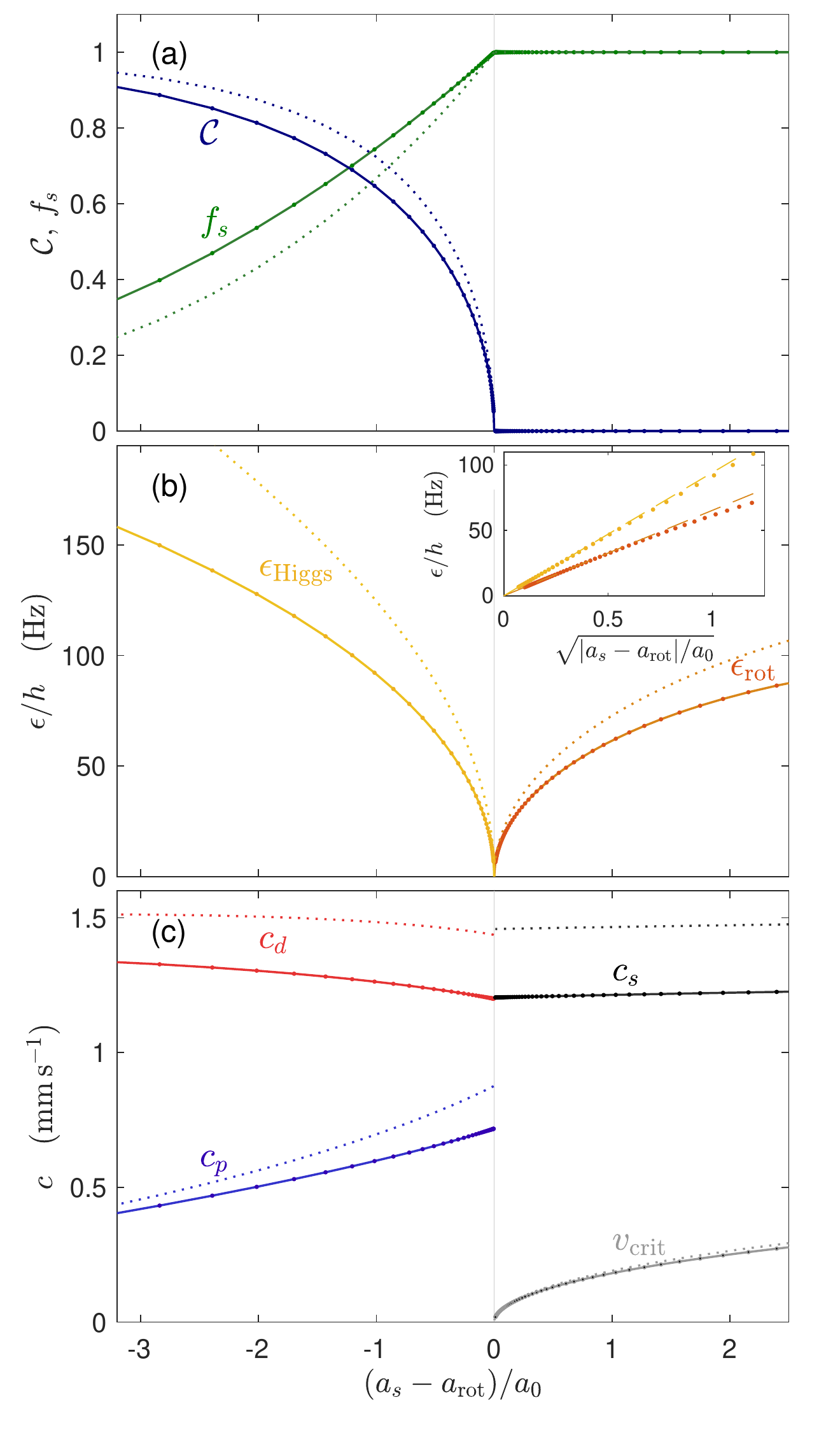}  
   \caption{Continuous transition between the uniform and modulated states at $n=2.5\times10^3\mu$m$^{-1}$ with $a_{\mathrm{rot}}=92.32\,a_0$. (a) Density contrast and superfluid fraction, (b) the energy of roton and Higgs mode, and (c) speeds of sound and critical velocity in the uniform state. The inset to (b) shows that the modes soften as $\sim\sqrt{|a_s-a_\mathrm{rot}|}$ in the approach to the transition, with the dashed lines being a linear fit to the data close to the transition. Small circles are results of our general 3D calculations and are fit with a solid line, except in the inset to (b) where these results are shown as small circles. For comparison the dotted lines show results of the reduced 3D theory where $a_{\mathrm{rot}}=90.56\,a_0$
   Other parameters as in Fig.~\ref{fig:dispersion}.
   \label{fig:sound}}
\end{figure}

 In Fig.~\ref{fig:sound} we consider characteristic properties of the system across the transition. Subplot (a) shows both the density contrast and the superfluid fraction. We can view the contrast as an order parameter for the crystalline order of the system. The finite superfluid fraction in the modulated state confirms that the system is in  a supersolid state. 
Note that here we have taken the superfluid fraction as the average of the upper  and lower bounds developed by Leggett and given by the expressions 
 \begin{align}
 f_s^+&=\frac{L}{n}\left[\int_\mathrm{uc}\frac{dz}{\int d\bm{\rho}\,|\psi|^2}\right]^{-1},\\
 f_s^-&=\frac{L}{n}\int d\bm{\rho}\left[\int_\mathrm{uc}\frac{dz}{|\psi|^2}\right]^{-1},
 \end{align}
  respectively \cite{Leggett1998a}.  These two bounds typically differ by less than a percent from each other and are in good agreement the superfluid fraction  obtained from the direct calculation of the nonclassical translational inertia of the system (see \cite{Smith2023a}). 
  
 We also consider the behavior of two characteristic excitations which reveal the approach to the transition from either side. In the uniform state the roton excitation plays this role and we define $\epsilon_{\mathrm{rot}}$ as the energy of the local minimum in the rotonic feature (e.g.~see \cite{Blakie2020a}). As $a_s$ approaches $a_\mathrm{rot}$ from above  $\epsilon_{\mathrm{rot}}\to0$. The softening of this mode leads to a dynamic instability causing the formation of spatial modulation \cite{Chomaz2018a,Hertkorn2021a}.
  In the modulated ground state a Higgs-like amplitude mode plays a key role in signifying the approach of the transition. The identification of Higgs-like mode is made with the $q_z\to0$ part of the third excitation band, because these excitations cause amplitude fluctuations of the crystalline order  \cite{Hertkorn2019a}. Here we define $\epsilon_{\mathrm{Higgs}} =\epsilon_{3,q_z=0}$ [see Fig.~\ref{fig:dispersion}(c1)], and as $a_s$ approaches $a_\mathrm{rot}$ from below  $\epsilon_{\mathrm{Higgs}}\to0$, and the modulated order disappears. The inset  to Fig.~\ref{fig:sound}(b) reveals that  the roton and Higgs energies both soften with an exponent of $\tfrac{1}{2}$, i.e.~$\epsilon\sim\sqrt{|a_s-a_\mathrm{rot}|}$, on their respective sides of the transition, consistent with the normal meanfield behaviour of the energy gap at a quantum phase transition \cite{Vojta2003a}.
  
Speeds of sound can be associated with the gapless excitation branches. The slope of the lowest band near $k_z=0$ in the uniform phase identifies the usual Bogoliubov sound for a BEC as  [see Fig.~\ref{fig:dispersion}(a)]
\begin{align} c_s=\frac{1}{\hbar}\left(\frac{\partial \epsilon_{1,k_z}}{\partial k_z}\right)_{k_z\to0}.
\end{align}  
Similarly, in  the modulated phase the two lowest bands can be used to define 
\begin{align}
c_p&=\frac{1}{\hbar}\left(\frac{\partial \epsilon_{1,q_z}}{\partial q_z}\right)_{q_z\to0},\\
 c_d&=\frac{1}{\hbar}\left(\frac{\partial \epsilon_{2,q_z}}{\partial q_z}\right)_{q_z\to0},
\end{align}
as the phase and density speeds of sound, respectively [see Fig.~\ref{fig:dispersion}(c1)]. Results for the speeds of sound are shown in Fig.~\ref{fig:sound}(c). The speeds are seen to change discontinuously across the transition, and is the basis of our identification of the transition as being second order\footnote{In Sec.~\ref{Sec:SFcompress} we see that this behavior appears as a discontinuity in the compressibility, which is a second derivative of the thermodynamic potential.}. For  the density we consider here the discontinuity is rather small with $c_d$ and $c_s$ being almost equal at the transition point [see Fig.~\ref{fig:compress}(c) for an example at a lower density where the discontinuity is larger]. 
The discontinuity in the speeds of sound arises from the avoided crossing between the low energy bands [e.g.~see Fig.~\ref{fig:dispersion}(b1)]. As we approach the transition from below the avoided crossing shifts towards $k_z\to0$ and hence affects the speeds of sound at the transition. 
We also note that the two speeds of sound in the stripe phase of spin-orbit  coupled BEC have been identified and studied in Refs.~\cite{Li2013a,Martone2021a}. 

We also show the critical velocity in the uniform state evaluated using the Landau criteria  
  \begin{align}
  v_\mathrm{crit}=\min_{k_z}\left(\frac{\epsilon_{1,k_z}}{\hbar k_z}\right).
  \end{align} 
For $a_s$ close to the transition the critical velocity is approximately given by $ v_\mathrm{crit}\approx\epsilon_\mathrm{rot}/\hbar k_\mathrm{rot}$ \cite{BECbook}, and the softening of the roton causes it to go to zero. The critical velocity and dynamics flow past an obstacle has been studied in soft-core models of supersolids \cite{Pomeau1994a,Kunimi2011a,Kunimi2012a}, but we have not made any extension of those ideas to our system.

While our main results in this paper are obtained by full numerical calculations, in Fig.~\ref{fig:sound} we also show the results of the  reduced 3D theory \cite{Blakie2020a,Blakie2020b,Ilg2023a}. The reduced theory makes a variational description of the transverse degrees of freedom, and reduces the calculation to an effective one-dimensional model.   In general the reduced theory produces qualitatively comparable results, although the speed of sound (particularly $c_s$ and $c_d$) tends to be significantly over estimated. This overestimation was also discussed in the context of the uniform state study presented in Ref.~\cite{Pal2020a}. These results suggest some caution is require in using reduced theory as a quantitative description of the system excitations.

\section{Structure factors and compressibility}\label{Sec:SFcompress}
 In addition to the spectrum of the excitations our interest here extends to the nature of the density fluctuations of the system, and their connection to compressibility. We can make this connection via the dynamical structure factor, which determines the response of the system to a density coupled probe, where the probe transfers momentum $\hbar\mathbf{k}$ and energy $\hbar\omega$ to the system. For the case of a momentum along the $z$-axis (i.e.~tube axis) and  the dynamic structure factor of the $T=0$ system is 
 \cite{Zambelli2000,Blakie2002a}
\begin{align}
S(k_z,\omega)&=\sum_{\nu}|\delta n_{\nu,k_z}|^2\delta\left(\hbar\omega-\hbar\omega_{\nu,q_z}\right),
\label{Eq:DSF}
\end{align}
where \begin{align}
\!\delta n_{\nu,k_z}=\int_\mathrm{uc}\!dz\int \!d\bm{\rho}\,[u^*_{\nu,q_z}(\mathbf{x})\!-\!v^*_{\nu,q_z}(\mathbf{x})]e^{ik_zz}\psi(\mathbf{x}).\label{dnkz}
\end{align} In this expression, and others where both $k_z$ and $q_z$ appear, the value of quasimomentum $q_z$ is fixed by $k_z$ reduced to the first Brillouin zone by an integer number of reciprocal lattice vectors $2\pi/L$ (also see \cite{Kramer2003a}).  
It is possible to measure the dynamic structure factor in cold-atom experiments using Bragg spectroscopy, which has been used to probe excitation properties  of dipolar BECs \cite{Bismut2012a,Petter2019a,Petter2021a}. Here our main interest lies in the static structure factor is given by
\begin{align}
S(k_z)\equiv\frac{\hbar}{nL}\int d\omega\,  S(k_z,\omega).
\end{align}
This can also be directly measured using high resolution \textit{in situ} imaging of the density fluctuations, e.g.~for the dipolar case see Refs.~\cite{Hertkorn2021a,Schmidt2021a}.
The static structure factor over a broad momentum range is shown in the insets of Figs.~\ref{fig:dispersion}(a2)-(d2). Here a divergence occurs at reciprocal lattice vector reflecting the periodic structure of the ground state\footnote{For  uniform state a finite peak occurs when the system has a roton, representing enhanced density fluctuations  \cite{Blakie2012a,Pal2020a}. This peak grows  as the roton softens, and diverges when the roton energy goes to zero.}.
We can examine the contribution from each band to the structure factor, i.e.~setting $S(k_z)=\sum_\nu S_\nu(k_z)$, where  $S_\nu(k_z)= \frac{1}{nL} |\delta n_{\nu,k_z}|^2$ is the contribution from the $\nu$-band. Results focusing on $S(k_z)$ and $S_\nu(k_z$) for low values of $k_z$  and $\nu=1,2,3$, are shown  (a2)-(d2). The contributions of higher bands $(\nu>3)$ are negligible. 
The avoided crossings in the spectrum (b1) are revealed in the behavior of the $S_\nu$ in (b2), where the weight smoothly transfers between the bands at the avoided crossings.
For the cases (c1,d1) where the lowest bands are separated, we see that the density band makes the dominant contribution to $S(k_z)$ in  the low-$k_z$ limit.  As $k_z$ increases the phase band contribution increases,  and  $|\delta n_{\nu,k_z}|$ diverges for both the phase and density bands as $k_z\to2\pi/L$. 

\begin{figure}[tbp]  
   \includegraphics[width=3.3in]{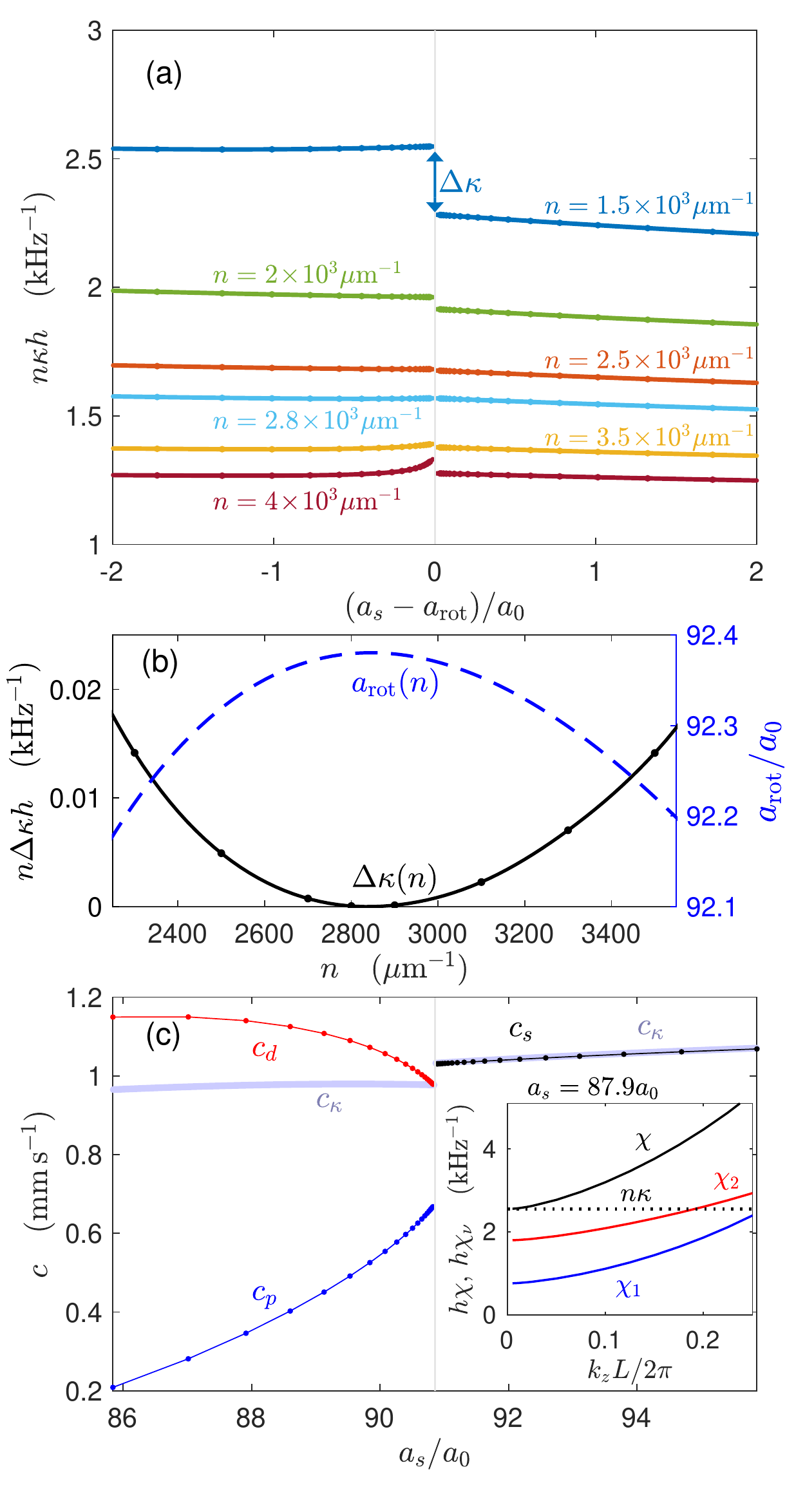}
   \caption{(a) Compressibility across the transition for various linear densities. (b) The magnitude of the compressibility jump at the transition point (black line, with markers, left axes) compared to $a_\mathrm{rot}$ (blue dashed line, right axes) as a function of density.   (c) For $n=1.5\times10^3 \mu$m$^{-1}$ we show the speeds of sound (line and markers), compared against  $c_\kappa$ (blue solid fine). Inset the contribution of the excitation branches to the compressibility sum rule for $a_s=87.9\,a_0$. Lowest (phase) excitation band (blue line), second density band (red line) and total of all bands (black line). Dotted horizontal line shows  $n\kappa$ for reference.
   Other parameters as in Fig.~\ref{fig:dispersion}. 
   \label{fig:compress}}
\end{figure}

The isothermal compressibility for the tube confined system can be defined as 
\begin{align}
\kappa=\frac{1}{n^2}\frac{\partial n}{\partial \mu},\label{Eq:compress}
\end{align}
and relates to the number fluctuations of the system in a large measurement cell (e.g.~see \cite{Klawunn2011,Bisset2013a,Baillie2014b}).
 We determine this directly from the ground state calculations by evaluating how the chemical potential changes with the average linear density.  This expression directly relates to the speed of sound in the uniform system as 
\begin{align}
n\kappa=\frac{1}{mc_s^2},\label{kappa_mc2}
\end{align}
 \cite{Pal2020a}. In Fig.~\ref{fig:compress}(a) we show results for the compressibility across the transition for systems of various densities. 
 For reference the parameter range for the data in this subplot is indicated in the phase diagram in Fig.~\ref{fig:schematic}(b).
 The compressibility exhibits a discontinuous jump of $\Delta\kappa=\kappa^--\kappa^+$ at the transition. Here $\kappa^\pm$ denotes the compressibility at the transition approaching from the below ($-$) or above ($+$). Results for $\Delta \kappa$ are presented in Fig.~\ref{fig:compress}(b), where we see that the discontinuity vanishes at $n\approx 2.85\times10^3 \mu$m$^{-1}$. 
We also show  $a_\mathrm{rot}(n)$ in this plot and observe that it is maximised around the same density value. The maximum in $a_\mathrm{rot}$ occurs from the competition between two-body interactions and the quantum fluctuation term \cite{Blakie2020a}. The simultaneous occurrence of these extrema suggests  that $\Delta \kappa$ vanishing is also related to this competition.

In Fig.~\ref{fig:compress}(c) we consider the speeds of sound for a lower density case than the results presented in Fig.~\ref{fig:sound}(b). For reference we have plotted the effective speed of sound 
\begin{align}c_\kappa=\frac{1}{\sqrt{mn\kappa}},
\end{align}
 obtained assuming relationship (\ref{kappa_mc2}) holds. We see that this value of $c_\kappa$ agrees with $c_s$ on the uniform side of the transition. In the modulated state it lies between the two speeds of sound, although at the transition its value is coincides with $c_d$. This indicates that at the transition the phase branch does not contribute to the long wavelength density fluctuations of the system.
 
 While the compressibility is calculated from of the ground state chemical potential (\ref{Eq:compress}), we can link it to the excitations via dynamical structure factor using the relation
\begin{align}
\int d\omega\,\frac{S(k_z,\omega)}{\omega}=\frac{1}{2}\chi(k_z),\label{compresssumrule}
\end{align}
where $\chi$ is the static density response function.
From Eq.~(\ref{Eq:DSF}) we see that $\chi(k_z) =\sum_\nu\chi_\nu(k_z)$, where
\begin{align} 
\chi_\nu(k_z)&\equiv \frac{2|\delta n_{\nu,k_z}|^2}{\epsilon_{\nu,q_z}}.
\end{align}
 In the long-wavelength limit the compressibility sum rule  \cite{stringari_1995,BECbook} is 
\begin{align}
\lim_{k_z\to0}\chi(k_z)=n\kappa. 
\end{align}
 We verify that this relationship holds in the 
inset to Fig.~\ref{fig:compress}(c). Similar to our treatment of the static structure factor in Figs.~\ref{fig:dispersion}(a2)-(d2), we can examine the contribution of each of the gapless bands to  (\ref{compresssumrule}), i.e.~$\chi_{\nu=1,2}$.  
The results, shown in the inset to Fig.~\ref{fig:compress}(c), reveal that in the modulated state the density and phase  bands both significantly contribute to $\chi$. The sum of these two saturates the contribution to $\chi$ for the range of $k_z$ values shown in the inset. In contrast our earlier results showed that for low $k_z$ the magnitude of $\delta n_{\nu,k_z}$ for the density band was significantly larger than the phase band\footnote{The density band does not saturate the f-sum rule \cite{BECbook}, but its contribution is much larger than the phase band.} [e.g.~see Figs.~\ref{fig:dispersion}(c2)-(d2)].  This behavior is different from the spin-orbit  stripe phase case where the density band excitations (propagating parallel to the crystal) exhaust the compressibility sum rule \cite{Martone2021a}.   The role of the phase band in increasing the compressibility relative to the density band was discussed in \cite{Ancilotto2013a} for a 3D soft-core supersolid, although no direct comparison of the elementary excitations to the compressibility was made.  
 
\section{Conclusions}

In this paper we have studied the ground states and excitations of a dipolar BEC as it transitions to a supersolid state in an infinite tube potential. We have focused on the regime where the $D=1$ crystalline order appears continuously, as characterized by the density contrast order parameter or the superfluid fraction. The compressibility and the speeds of sound obtained from the gapless energy bands are generally discontinuous across the transition, consistent with the transition being second order. 

It is interesting to consider the prospects for measuring our predictions in experiments. Recent experiments with optical lattices have measured the speed of sound and determined the superfluid fraction using Bragg spectroscopy \cite{Tao2023a} and collective mode excitation \cite{Chauveau2023a}. These techniques could be applied to the dipolar system noting that Bragg spectroscopy has been used to measure the anisotropy of sound in a dipolar BEC \cite{Bismut2012a}, and to probe the free-particle excitations of a dipolar supersolid \cite{Petter2021a}.   Also, various forms of collective mode spectroscopy have already been demonstrated in this system \cite{Tanzi2019b,Guo2019a,Natale2019a,Hertkorn2021a}.
The compressibility relates to the  number fluctuations in a large measurement cell \cite{Klawunn2011,Bisset2013a,Baillie2014b} and has been measured experimentally in ultra-cold atoms experiments using \textit{in-situ} density and density fluctuation measurements  \cite{Gemelke2009a,Hung2011,Ku2012a,Ye-Ryoung2012a} (also see \cite{Poveda-Cuevas2015a}). Related measurements have been performed in dipolar BECs and supersolids to determined the static structure factor \cite{Hertkorn2021a,Schmidt2021a} and could be extended to probe compressibility.

\section*{Acknowledgments}
\noindent PBB  acknowledges  use of New Zealand eScience Infrastructure (NeSI) high performance computing facilities. PBB and DB acknowledge support from the Marsden Fund of the Royal Society of New Zealand.
LC acknowledges support from the European Research Council (ERC) under the European Union’s Horizon Europe research and innovation program under grant number 101040688 (project 2DDip), and from the Deutsche Forschungsgemeinschaft (DFG, German Research Foundation) through project-ID 273811115 (SFB1225 ISOQUANT) and under Germany’s Excellence Strategy EXC2181/1-390900948 (the Heidelberg Excellence Cluster STRUCTURES). Views and opinions expressed are however those of the authors only and do not necessarily reflect those of the European Union or the European Research Council.
 
%

\end{document}